\documentclass[10pt]{article}

\usepackage{amsmath}
\usepackage{amssymb}

\usepackage{graphicx}

\usepackage{cite}

\usepackage{color} 

\usepackage[labelfont=bf,labelsep=period,justification=raggedright]{caption}

\makeatletter
\renewcommand{\@biblabel}[1]{\quad#1.}
\makeatother

\date{}

\begin{document}

\begin{flushleft}
{\Large
\textbf{Evolutionary accessibility of mutational pathways}
}
\\
Jasper Franke$^{1}$, 
Alexander Kl\"ozer$^{1}$, 
J. Arjan G.M. de Visser$^{2}$ and Joachim Krug$^{1,\ast}$
\\
\bf{1} Institute of Theoretical Physics, University of Cologne, K\"oln, Germany
\\
\bf{2} Laboratory of Genetics, Wageningen University, Wageningen, Netherlands
\\
$\ast$ E-mail: krug@thp.uni-koeln.de
\end{flushleft}

\section*{Abstract} 
Functional effects of different mutations are known to combine to the total effect in highly nontrivial ways. For the trait under evolutionary selection (`fitness'), measured values over all possible combinations of a set of mutations yield a fitness landscape that determines which mutational states can be reached from a given initial genotype. Understanding the accessibility properties of fitness landscapes is conceptually important in answering questions about the predictability and repeatability of evolutionary adaptation. 
Here we theoretically investigate accessibility of the globally optimal state on a wide variety of model landscapes, including landscapes with tunable ruggedness as well as neutral `holey' landscapes. We define a mutational pathway to be accessible if it contains the minimal number of 
mutations required to reach the target genotype, and if fitness increases in each mutational step. 
Under this definition accessibility is high, in the sense that at least one accessible pathway
exists with a substantial probability that approaches unity as the dimensionality of the fitness landscape (set by the number of mutational loci) becomes large. At the same time the number of alternative accessible pathways grows without bound. We test the model predictions against an empirical 8-locus fitness landscape obtained for the filamentous fungus \textit{Aspergillus niger}. By analyzing subgraphs of the full landscape containing different subsets of mutations, we are able to probe the mutational distance scale in the empirical data. 
The predicted effect of high accessibility is supported by the empirical data and very robust, which we argue to reflect the generic topology of sequence spaces. Together with the restrictive assumptions that lie in our definition of accessibility, this implies that the globally optimal configuration should be accessible to genome wide evolution, but the repeatability of evolutionary trajectories is limited owing to the presence of a large number of alternative mutational pathways.

\section*{Author Summary}
Fitness landscapes describe the fitness of related genotypes in a given environment, and can be used to identify which mutational steps lead towards higher fitness under
particular evolutionary scenario's. The structure of a fitness landscape results from the way mutations interact in determining fitness, and can be smooth when mutations
have multiplicative effect or rugged when interactions are strong and of opposite sign.  Little is known about the structure of real fitness landscapes.  Here, we study
the evolutionary accessibility of fitness landscapes by using various landscape models with tuneable ruggedness, and compare the results with an empirical fitness
landscape involving eight marker mutations in the fungus \textit{Aspergillus niger}. We ask how many mutational pathways from a low-fitness to the globally optimal genotype are
accessible by natural selection in the sense that each step increases fitness.  We find that for all landscapes with lower than maximal ruggedness the number of
accessible pathways increases with increases of the number of loci involved, despite decreases in the accessibility for each pathway individually.  We also find that
models with intermediate ruggedness describe the \textit{A. niger} data best.

\section*{Introduction}

Mutations are the main sources of evolutionary novelty, and as such
constitute a key driving force in evolution. They 
act on the genetic constitution of an organism at very different levels, from 
single nucleotide substitutions to large-scale chromosomal modifications.
Selection, a second major evolutionary force, 
favors organisms best adapted to
their respective surroundings. Selection
acts on the fitness of the organism. How fitness is connected
to specific traits such as reproduction or survival  
depends strongly on the environmental 
conditions, but indirectly it can be viewed
as a function of the organism's genotype.

If one considers mutations at more than one locus, it is not at all
clear how they combine in their final effect on fitness. Two
mutations that individually have no significant effect on a trait
under selection can in combination
be highly advantageous or deleterious. Well known examples for such
epistatic interactions\cite{epistasis} include resistance evolution in 
pathogens\cite{Hall2002,wddh2006,L2009} or metabolic 
changes in yeast\cite{sdck2005}.  
In general, the presence of epistatic
interactions makes the fitness landscape 
more rugged, particularly when epistasis affects the
sign of the fitness effects of mutations \cite{Weinreich2005b,pkwt2007,Kvitek2011}.
Fitness landscapes are most easily dealt with in the context of asexual haploid organisms, and
we will restrict our considerations here to this case.

In a remarkable recent development,
several experimental studies have probed the effect 
of epistatic interactions on fitness 
landscapes\cite{wddh2006,pkwt2007,L2009,Lunzer2005,pdk2009,Kogenaru2009,Dawid2010,DaSilva2010,Tan2011,Chou2011,Khan2011}. 
Most of these studies are based
on two genotypes, one that is well adapted to the given environment,
and another that differs by a known, small set of mutations;
the largest landscapes studied so far involve five mutations \cite{wddh2006,pdk2009,Khan2011}. All
(or some fraction of the) intermediate genotypes are then constructed and their 
fitness measured. However, selection in natural populations does not
act on small, carefully selected sets of mutations, 
but rather on all possible beneficial mutations that occur anywhere in the genome,
making the number of possible mutations many orders of
magnitude greater than those considered in empirical
studies. 

\begin{figure}[!ht]
\begin{center}
\includegraphics[width=0.3\textwidth]{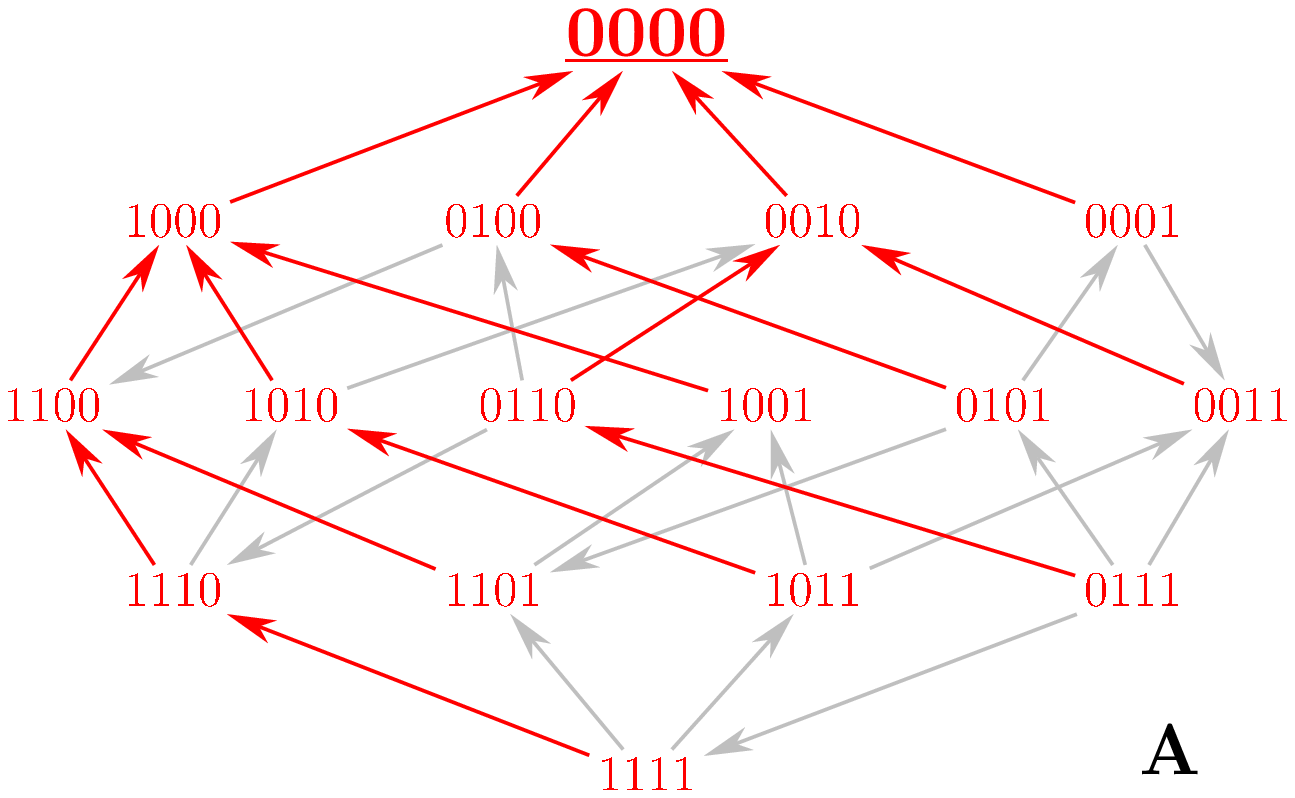} \hspace*{0.2cm}\includegraphics[width=0.3\textwidth]{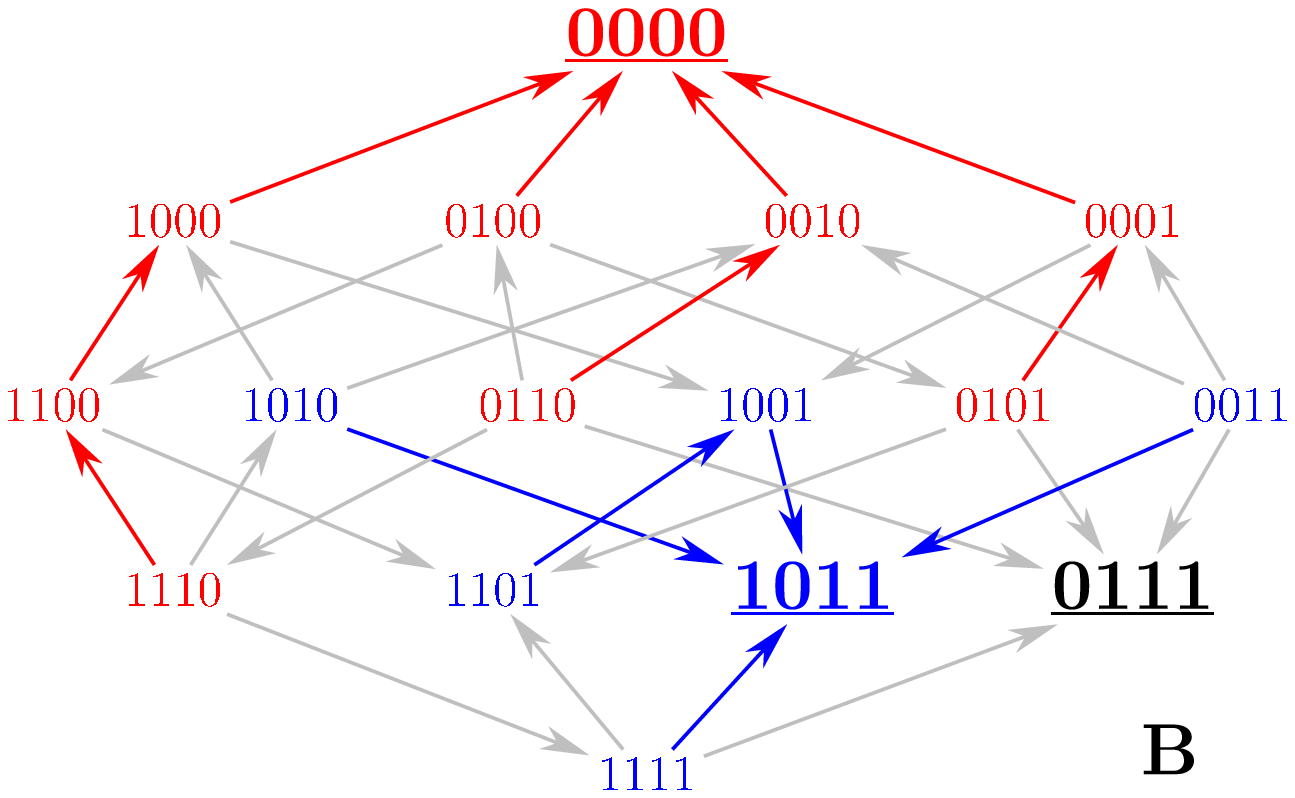}\hspace*{0.2cm}\includegraphics[width=0.33\textwidth]{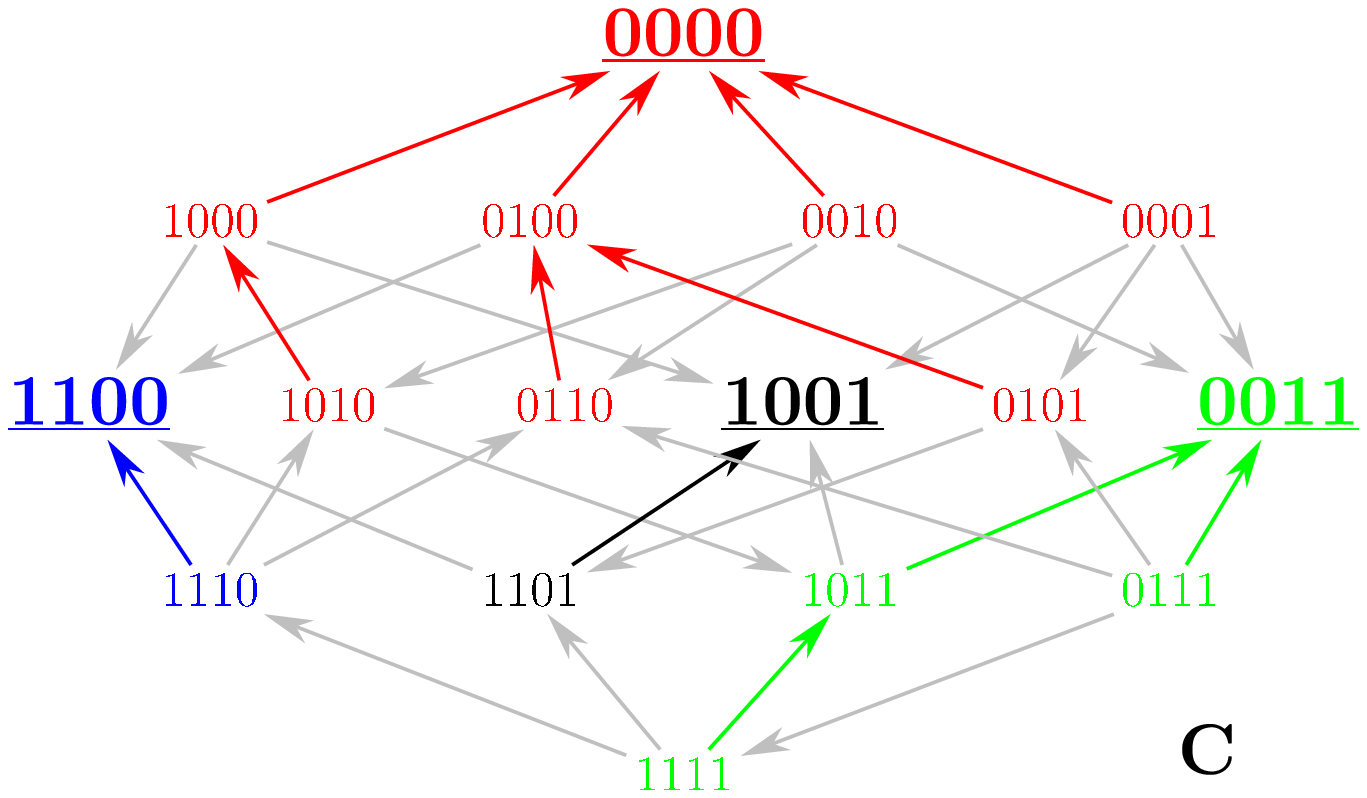}
\end{center}
\caption{{\bf Graphical representation of three fitness landscapes of size $m=4$ extracted from the empirical
  8-locus fitness data set for \emph{A. niger}}. The presence/absence of a given mutation is indicated by 1/0. Arrows point towards higher
  fitness, local maxima are enlarged and underlined, and colors mark basins of
  attraction of maxima under a greedy (steepest ascent) adaptive walk. (A) All combinations of mutations \emph{argH12, pyrA5, leuA1, oliC2}. This landscape has a single fitness maximum (the wildtype), but only 9 out of 4!=24 paths from \{1111\} to \{0000\} are accessible.
  (B) Mutations \emph{argH12, pyrA5, leuA1, pheA1}. This landscape has three maxima and no accessible path. (C) Mutations \emph{fwnA1, leuA1, oliC2, crnB12}. The landscape has four maxima and 2 accessible paths.}
\label{fl}  
\end{figure}

Figure \ref{fl} shows three sample landscapes obtained from an empirical
8-locus dataset of fitness values for the fungus \textit{Aspergillus niger} 
originally obtained in \cite{dhv1997} (see \textbf{Materials and Methods} for details on the
data set and its representations). These landscapes display a wide variation in 
topography, and despite their moderate size of $2^4=16$ genotypes, the combinatorial
proliferation of possible mutational pathways 
makes it difficult to infer the adaptive fate of a population
without explicit simulation\cite{pdk2009}. 
In fact, in view of the broad range of possible landscape topographies, 
even a thorough understanding  of evolution on one of
these landscapes would be of limited use when confronted with
another subset of mutations or even fitness landscapes from a different
organism. Instead, one would like to understand and quantify the typical features
of \textit{ensembles} of fitness landscape, where an ensemble can be formed e.g.
by selecting different subsets of mutations from an empirical data set, or by 
generating different realizations of a random landscape model.  

Although genome-wide surveys of pairwise epistatic interactions have recently become
feasible \cite{Costanzo2010}, exploring an entire fitness landscape on a genome-wide
scale remains an elusive goal. In this situation theoretical
considerations are indispensable to assess the influence of epistasis on
the outcome of evolutionary adaptation. Here, we aim to perform part
of this task by answering  the following question: Does epistasis
make the global fitness optimum selectively inaccessible?

This question has a long history in evolutionary theory, and two contradictory intuitions 
can be discerned in the still ongoing debate \cite{epistasis}. 
One viewpoint generally attributed to Fisher \cite{Fisher} emphasizes the proliferation of 
mutational pathways in high dimensional genotype spaces to argue that, because of the sheer
number of possible paths, accessibility will remain high. The second line of argument originally
formulated by Wright \cite{Wright1932}, 
and more recently promoted by Kauffman \cite{kauffman} and others, focuses instead on the
proliferation of local fitness maxima, which present obstacles to adaptation and reduce 
accessibility with increasing genotypic dimensionality. Here we
show that both views are valid at a qualitative level, but that Fisher's scenario prevails on the basis of
a specific, quantitative definition of accessibility, since the number of accessible pathways grows much faster with landscape
dimensionality than the inaccessibility per pathway 
as long as the fitness landscape is not completely uncorrelated.
Moreover, 
our analysis of accessibility in the empirical \textit{A. niger} data set illustrated
in Fig.\ref{fl} shows how evolutionary accessibility can be used to quantify the degree of sign
epistasis in a given fitness landscape.

\subsection*{Mathematical framework}

The dynamics of adaptation of a haploid asexual population on a given 
fitness landscape is governed by
population size $N$, selection strength $s$ and mutation rate $u$, and  
different regimes for these parameters have been identified
\cite{gillespie, hartl_clark,Jain2007}. Here we assume a
`strong-selection/weak mutation' (SSWM)
regime\cite{Gillespie1983,Orr2002}, which implies that mutations are
selected one by one and prohibits
the populations from crossing valleys of
fitness. In natural populations of sufficient size, a number of double mutants is
present at all times, 
and the crossing of fitness valleys can be 
relatively facile\cite{Weinreich2005,Weissman2009}; the SSWM assumption
may therefore seem overly restrictive. However, we
will see that even under these conditions, the landscapes considered
are typically very accessible.

In the remainder of the paper, the genetic configuration of the
organism will be represented as a binary sequence
$\boldsymbol{\sigma}=\{\sigma_1, \dots, \sigma_L\}$ of total length
$L$, 
where $\sigma_i=1$ ($\sigma_i=0$) stands for the presence (absence)  
of a given mutation in the landscape of interest.
The SSWM assumption together with the fact that we only consider
binary sequences gives the configuration space the topological
structure of a hypercube of dimension $L$.
Accessibility can then be quantified by studying the \textit{accessible mutational
paths} \cite{Kauffman1987,Hall2002,wddh2006}. 
A mutational path is a collection of point
mutations connecting an initial state $\boldsymbol{\sigma}_{I}$ with a
final state $\boldsymbol{\sigma}_F$. If these two states differ at $l$
sites, there are $l!$ shortest paths connecting them, corresponding to the different
orders in which the mutations can be introduced into the population \cite{Gokhale2009}. 
The assumed weak mutation rate implies that paths longer
than the shortest possible path have a much lower probability of occurrence
and hence are not considered here, adding to the constraints already
imposed on accessibility.
A mutational path is considered selectively accessible (or accessible
for short) if the fitness values encountered along it are monotonically
increasing; thus along such a path, the population never encounters a
decline in fitness. If two states are separated by a fitness valley,
the path is inaccessible. Neutral mutations are generally not detected in the empirical
fitness data sets of interest here, though they may be present at a finer scale of resolution
\cite{Wagner2008}. In our modeling we therefore assume that the fitness values of neighboring
genotypes can always be distinguished (but see the discussion of the holey landscape model below). 

Unlike Ref.\cite{wddh2006} we only consider
whether a given path is 
at all accessible or not, independent of the probability of the path actually
being found by the population. Our reason for focusing on this
restricted notion of accessibility is that it can be formulated solely
with reference to the underlying fitness landscape, without the need to
specify the adaptive dynamics of the population (see also \textbf{Discussion}).
The endpoint of the paths considered here, much like in the
experimental studies\cite{wddh2006,L2009,pdk2009}, is the global fitness
maximum, and the starting point is the 
'antipodal' sequence which differs from
the optimal sequence at all $L$ loci. 
 Because it is at the
opposite end of the configuration space, these are the longest direct paths. 
As such, they are \textit{a priori} the least likely to be
accessible 
and thus give a lower limit on the accessibility of
typical paths (note that the mean length of the path from a randomly chosen genotype
to the global maximum is $L/2$). 

For a fitness landscape comprised of up to $L$ mutations, there are
a total of $L!$ paths connecting the
antipodal sequence to the global maximum. How many of them are
selectively accessible in the sense described above? Given
that natural selection is expected to act genome-wide, we are
interested in the behavior of accessibility properties when the number
of loci $L$ becomes very large. Two questions are of particular interest: 
What is the probability of finding at least one accessible path, and 
what number of accessible paths can one expect to find on average? 
The first question addresses the overall accessibility of the global fitness
maximum\cite{ch2010}, while the second question is relevant for the repeatability
of evolution: If there are many possible mutational pathways connecting the initial
genotype to the global maximum, depending on population dynamics 
different pathways can be chosen in 
replicate experiments and repeatability will be low. 
To address these questions in a quantitative way, consider a 
sample of fitness landscapes, obtained e.g. as random realizations
of a landscape model or by choosing subsets of mutations from a large
empirical data set (see Fig.~\ref{fl}). The fraction of these that
have exactly $n$ accessible paths is denoted by $p_{L}(n)$, 
and gives an estimate of the probability
that a given fitness landscape has $n$ accessible paths
(cf. Fig.~\ref{pnl}). The expected number of paths is given by   
the mean of this probability distribution,
\begin{equation}
\label{mean}
\left<n_{L}\right>=\sum_{n=0}^{L!}np_{L}(n), 
\end{equation}
and $1-p_{L}(0)$
is the probability to find at least one accessible path. The behavior of these two 
quantitities will be investigated in the following, both for model landscapes and on the 
basis of empirical data.

\section*{Results}

\subsection*{House of Cards (HoC) model}

\begin{figure}[!ht]
\begin{center}
\includegraphics[scale=0.6]{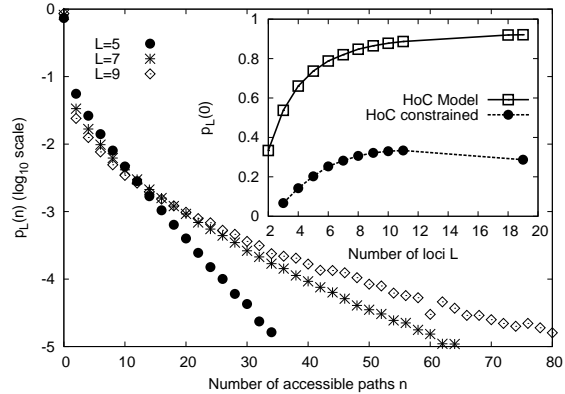}
\end{center}
  \caption{{\bf Accessibility of mutational pathways in the House-of-Cards model.} 
Main figure shows the distribution of the number of accessible
    paths for three different sequence lengths in the HoC model in 
    semi-logarithmic scales. The value of $p_{L}(0)$ is an outlier,
    indicating that a large fraction of landscapes have no accessible
    paths at all. This is a typical feature of rugged fitness landscapes of moderate dimensionality
$L$, see Figs.~S4 and S5. Inset shows  
    $p_{L}(0)$ as function of L for the HoC model. The top
    curve makes no assumptions about the antipodal sequence, while the
    bottom curve assumes it to be the global fitness minimum. Note the
   decline in the bottom curve.}
\label{pnl} 
\end{figure}

Consider a model where fitness
values are uncorrelated and a single mutation may change fitness
completely\cite{Jain2007,kauffman,Kauffman1987}; following Kingman\cite{k1978} we
refer to this as the `House of Cards' model.
In real organisms one expects fitnesses of closely related genotypes
to be at least somewhat correlated, and in this sense the HoC model
serves as a null model. 
The expected number of accessible
paths can be computed exactly by a simple order statistics
argument\cite{fwk2010}. Each of the $L!$ shortest paths contains
$L+1$ genotypes. Out of the $L+1$ fitness values encountered along a
path, all but the last one (which is known to be the global maximum) 
are arranged in any order with equal probability. One of the $L!$ 
possible orderings is monotonic in fitness, hence for the HoC model
\begin{equation}
\label{meanHoC}
\langle n_{L}\rangle=\frac{L!}{L!}=1
\end{equation}
for all $L$. The probability $p_{L}(0)$
of not finding any path is more difficult to compute and was so far 
only analyzed by numerical simulations. 
We find that for sequence lengths up to 
$L=20$, $p_{L}(0)$ appears to approach unity, see inset of Fig.
\ref{pnl} and Fig.~S1. Whether this is asymptotically true remains to be established,
but the scaling plot in the inset of Fig.~\ref{rmf_pl0} suggests that $p_{L}(0)$ is indeed monotonically
increasing for all finite $L$.

This behavior changes drastically when the antipodal
state is required to be the global fitness minimum. This case
was considered previously by Carneiro and Hartl\cite{ch2010}, who postulated 
that $p_L(0)$ saturates to an asymptotic value around $1/3$ for large
$L$. However by continuing the simulations to $L=19$, one sees a clear
decline (inset of Fig.~\ref{pnl}), indicating that accessibility
\textit{increases} with increasing $L$. We will see in the following
that this is in fact the generic situation. 

\subsection*{Rough Mount Fuji (RMF) model}

Next we ask what happens if some fitness
correlations are introduced. The Rough Mount Fuji (RMF) model\cite{aEtAl2000}
accomplishes just that: Denoting the number of mutations separating a
given genotype $\boldsymbol{\sigma}$ from the global optimum
by $d_{\boldsymbol{\sigma}}$, the RMF
model assigns fitness values according to
\begin{equation}
\label{RMFfitness}
F_{\boldsymbol{\sigma}}=-\theta \cdot d_{\boldsymbol{\sigma}} +
x_{\boldsymbol{\sigma}}, 
\end{equation}
where $\theta\geq 0$ is a constant and the
$x_{\boldsymbol{\sigma}}$ 
are independent normal random variables with zero mean and unit variance. 
When $\theta\equiv 0$ the RMF reduces to the HoC case, and thus
it can serve as starting point for approximate calculations to first
order in $\theta$. For the expected number of accessible paths one
obtains\cite{fwk2010} 
\begin{equation}
\label{meanRMF}
\langle n_{L} \rangle \approx 1+\theta L(L-1)\gamma,
\end{equation}
where $\gamma > 0$ and terms of higher order have been neglected (see also 
Eq.~(\ref{pi})). In this limit $\langle n_{L} \rangle$ grows like $L^2$ for large $L$ and 
constant $\theta$. Compared to the HoC case $\theta=0$, this shows
that the large $L$-behavior of a landscape with even the slightest 
correlation between fitness values is substantially different from
the case without correlations. 

\begin{figure}[!ht]
\textbf{A} \hspace*{8.cm} \textbf{B}
\begin{center}
  \includegraphics[scale=0.6]{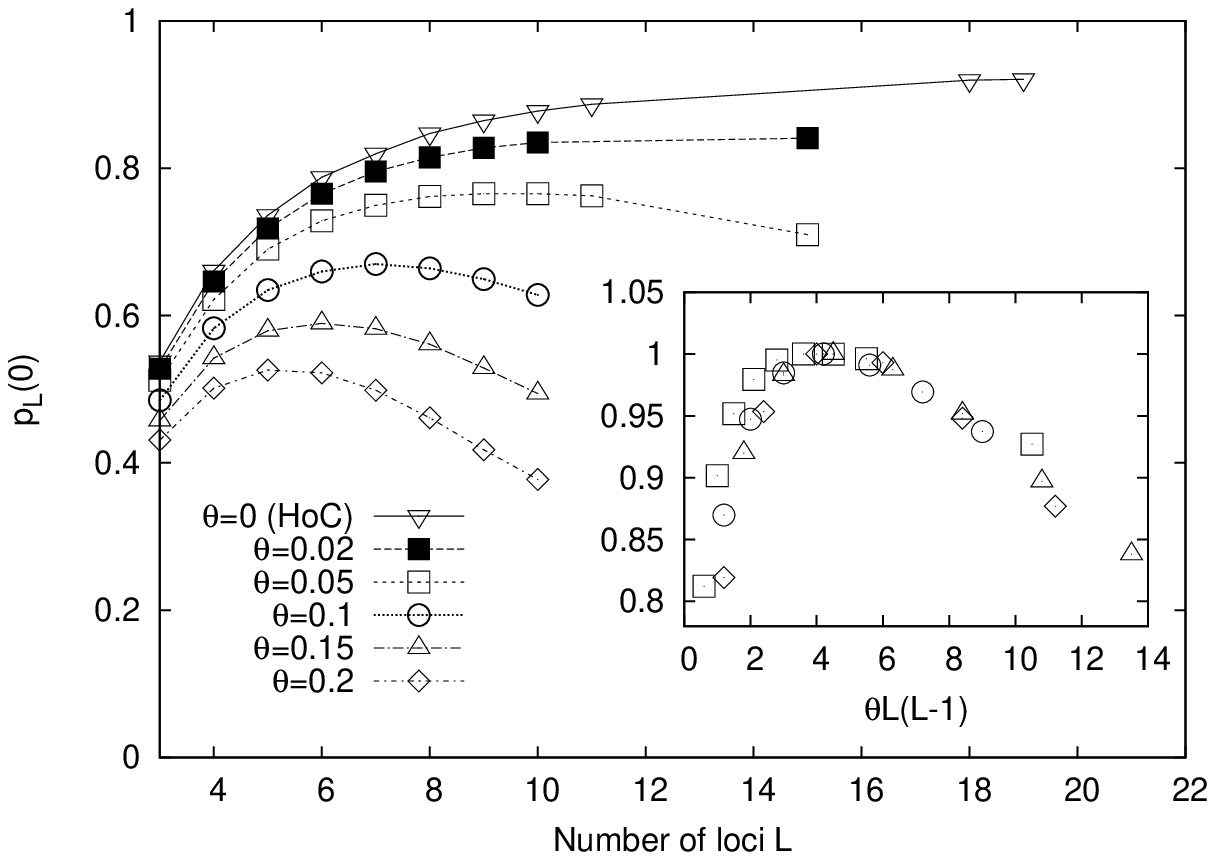}\hspace*{0.5cm}\includegraphics[scale=0.6]{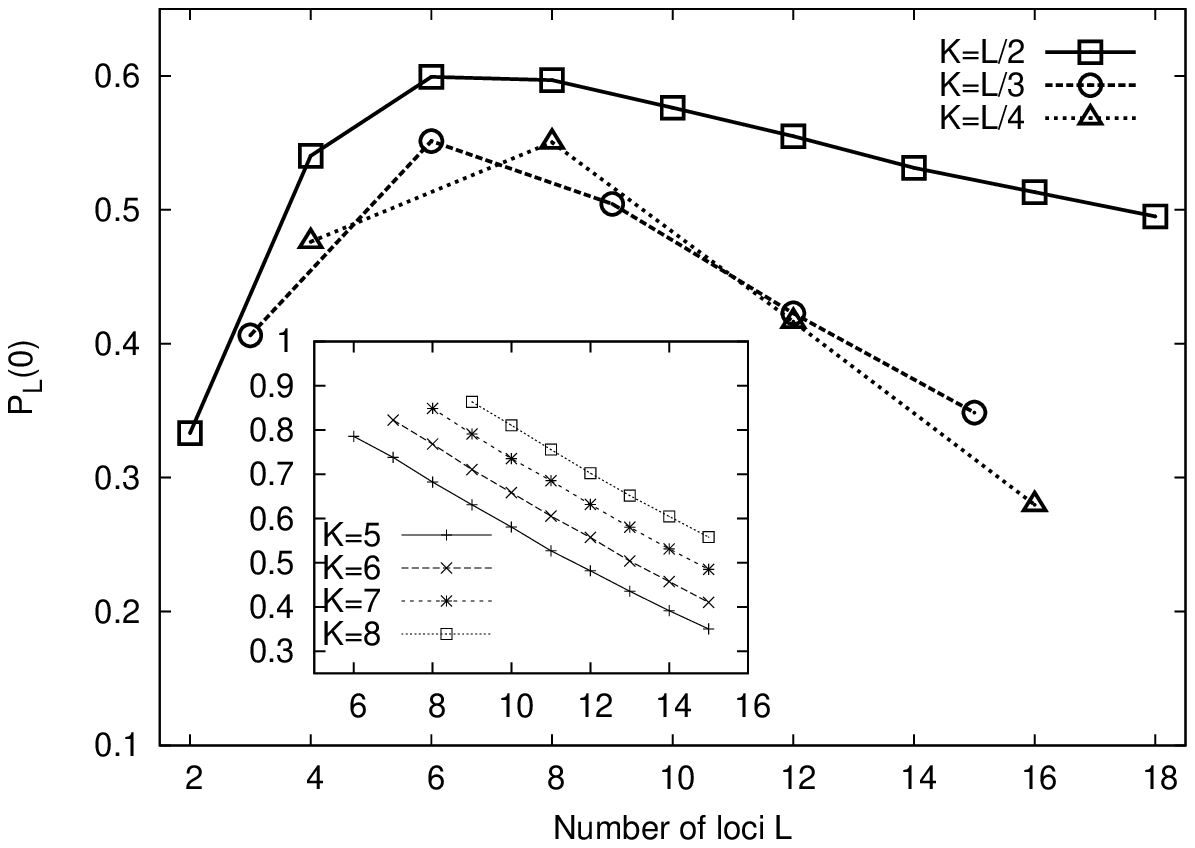}
\end{center}
\caption{{\bf Accessibility in fitness landscape models with tunable ruggedness.} (A) Behavior of 
  $p_{L}(0)$ in the RMF model as function of the correlation parameter
  $\theta$. Inset shows normalized rescaled curves, all taking their
  maximum at $\theta L(L-1)\sim 4$. This implies that $p_L(0)$ increases monotonically only for
  $\theta \equiv 0$. (B) Probability $p_L(0)$ for the $LK$ model as a function of $L$ at fixed $K/L$ (main figure) and fixed $K$ (inset), respectively.} 
\label{rmf_pl0}
\end{figure}

The probability of finding no accessible paths was again obtained by numerical simulation,
and is shown in Fig.~\ref{rmf_pl0}(a). In striking contrast to the unconstrained
HoC model, the probability $1-p_L(0)$ of finding at least one accessible path 
is seen to increase for large $L$. Motivated by the result (\ref{meanRMF}), in the inset
of Fig.~\ref{rmf_pl0}(a)
the simulation results are plotted as a function of $\theta L(L-1)$, which leads to an approximate
collapse of the different data sets.  
On the basis of these results we conjecture that, for any $\theta>0$, the probability
$p_L(0)$ decreases for large $L$, and most likely
approaches zero asymptotically for $L \to \infty$.

\subsection*{LK model}

Better known as the NK-model\cite{kauffman,Kauffman1989},
this classical model explicitly takes into account epistatic interactions
among different loci. Each of the $L$ sites in the
genome is assigned a certain number $K$ of other sites with which it 
interacts, and for each of the possible $2^{K+1}$ states of this
set of interacting loci the site under
consideration contributes to the fitness by a random amount. Thus the
parameter $0\leq K \leq L-1$ defines the size of the epistatically
interacting parts of the sequence and provides a measure for the amount
of epistasis. Like the RMF model, the $LK$ model reduces in one limit
to the HoC case,  
which is realized for
$K=L-1$. 

Due to the construction of
the model, even local properties such as the number of
local fitness optima \cite{dl2003, lp2004} are generally very difficult to
compute. Figure \ref{rmf_pl0}(b) shows the variation of $p_L(0)$
with $L$ obtained from numerical simulations of the LK-model. 
In this figure two different relations between $K$ (the number of interacting loci) and $L$
(the total number of loci) were employed. 
In the main plot the fraction of interacting loci $K/L$ was kept constant. 
Under this scenario, the curves show a non-monotonic
behavior of $p_L(0)$ similar to that of the RMF model at constant epistasis parameter
$\theta$. In the inset, the number of interacting loci $K$ is kept fixed, which results
in a monotonic decrease of $p_L(0)$. A third possibility is to 
fix the \emph{difference} $L-K$ (the number of non-interacting loci), see Fig.~S2. In this case one
can argue that for $L \gg 1$, the difference in behavior between
$K=L-1$ and $K=L-2$, say, should not be substantial, and indeed
the curves for $p_{L}(0)$ seem to be monotonically increasing with $L$,
showing qualitatively the same behavior as the curve for $K=L-1$, which is equivalent to the
HoC model. Finally, 
in Fig.~S3 we show the expected number of
accessible paths for different values of $K$
and $L$. The data are seen to interpolate smoothly between the known
limits $\langle n_L \rangle = L!$ for $K = 0$ and $\langle n_L \rangle =
1$ for $K = L-1$.

\subsection*{Holey landscapes}

The neutral theory of evolution\cite{kimura} 
implies a very simple, flat fitness landscape without
maxima or minima. When strongly deleterious mutations are included, 
the resulting fitness landscape has plateaus of viable states
and stretches of lethal states\cite{MS1970}. Such 'holey' landscapes can be
mapped\cite{Gavrilets2004}  
to the problem of percolation, a paradigm of statistical physics\cite{Stauffer}.
In percolation, each configuration is
either viable (fitness $1$) with probability $p$ or lethal (fitness
$0$) with probability $1-p$, independent of the others. 
Our definition of accessibility must be adapted
in this case, as there is no notion of increasing fitness and no
global fitness optimum. However, one can still ask the
question whether it is possible to get from one end of configuration space
to the other on a shortest path of length $L!$ without encountering a
`hole', i.e. a non-viable state. Apart from the restriction to shortest paths,
the quantity $1-p_L(0)$ of finding at least one connecting path then corresponds
to the percolation probability. 

The percolation problem on the hypercube differs from the standard 
case of percolation on finite-dimensional lattices\cite{Stauffer} in that 
the parameter $L$ represents both the dimensionality and the diameter
of the configuration space. 
Percolation properties are therefore
described by statements that hold asymptotically for large $L$ under some
suitable scaling of the viability probability $p$ \cite{GG1997,r1997}. Specifically, when
$p = \lambda/L$ for some constant $\lambda$, it is known that for
$\lambda > 1$ a giant connected set  of viable genotypes emerges for $L \to
\infty$. Conversely, taking $L \to \infty$ at fixed $p$ 
one expects that two antipodal genotypes are connected by
a path with a probability approaching unity. 
Indeed, the simulation results shown in Fig.~S4
support the conjecture that the quantity corresponding to
$p_L(0)$ vanishes for large $L$ and any $p > 0$. 
The equivalent of computing $\langle n_{L} \rangle$ is straightforward: 
The probability that $L$ consecutive states 
are viable factorizes by independence of the fitness values to the
product of the individual probabilities of viability, to simply yield
$p^L$, which, as $p<1$, decays exponentially. We already know that
there are $L!$ possible paths in the sequence space, thus we find
\begin{equation}\label{percon}
  \langle n_{L}\rangle =p^LL!.
\end{equation}
Since $L!$ grows faster than $p^L$ declines, $\langle n_{L} \rangle$
grows without bounds for large $L$.

\subsection*{Comparison to empirical data}

Next we compare the predictions of the models described so far
to the results of the analysis
of a large empirical data set obtained from fitness measurements
for the asexual filamentous fungus \textit{A. niger}. 
As described in more detail in \textbf{Materials and Methods}, we analyzed
the accessibility properties of ensembles of subgraphs containing
subsets of $m=2,..,6$ out of a total of 8 mutations which are
individually deleterious but display significant epistatic
interactions \cite{dhv1997}. 
The full data set contains fitness values for 186 out of the $2^8 = 256$ possible 
strains, and statistical analysis shows that the 70 missing
combinations can be treated as non-viable genotypes with zero fitness.
The distribution of the non-viable genotypes in the subgraph ensemble 
is well described by a simple two-parameter model 
which reveals that the lysine deficiency mutation \textit{lysD25} is about 25 times more likely
to cause lethality than the other seven mutations (see \textbf{Materials and Methods}).

\begin{figure}[!ht]
\textbf{A} \hspace*{8.cm} \textbf{B}
\begin{center}
\includegraphics[scale=0.6]{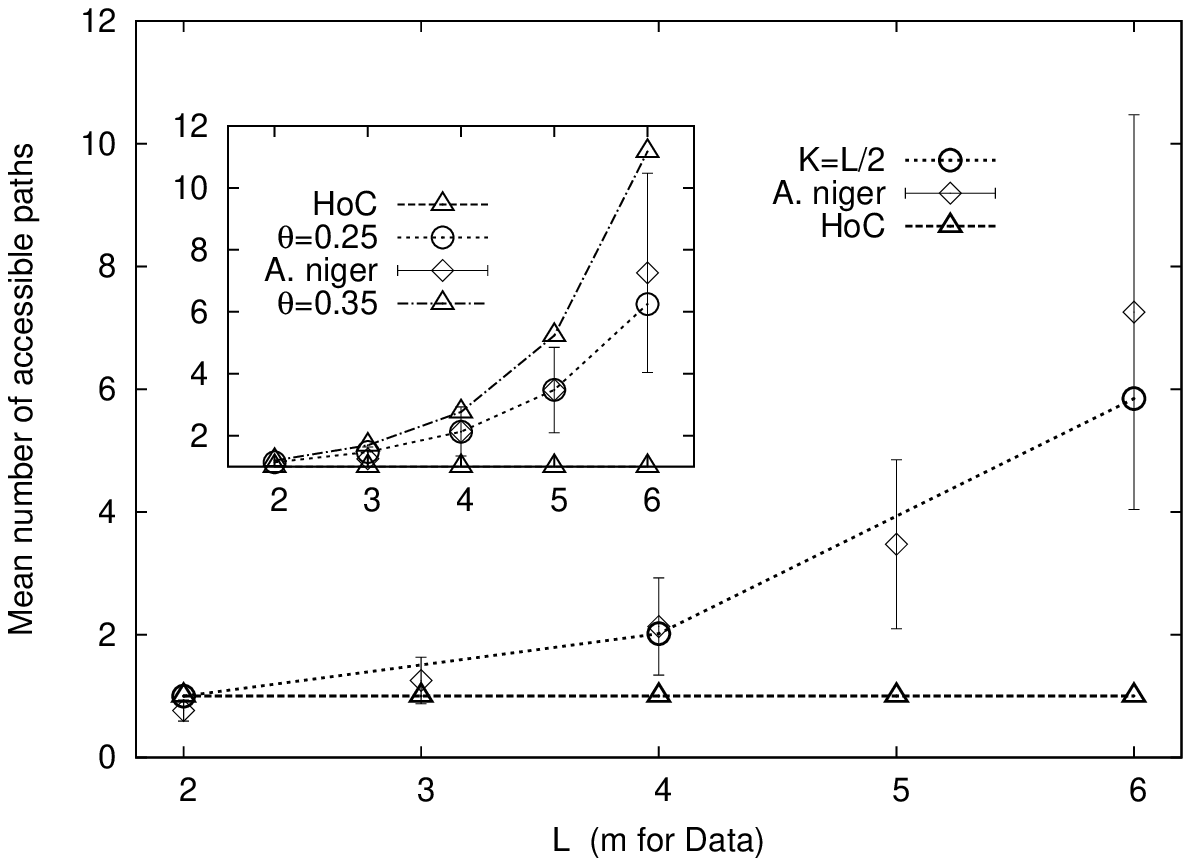} \hspace*{0.5cm} \includegraphics[scale=0.6]{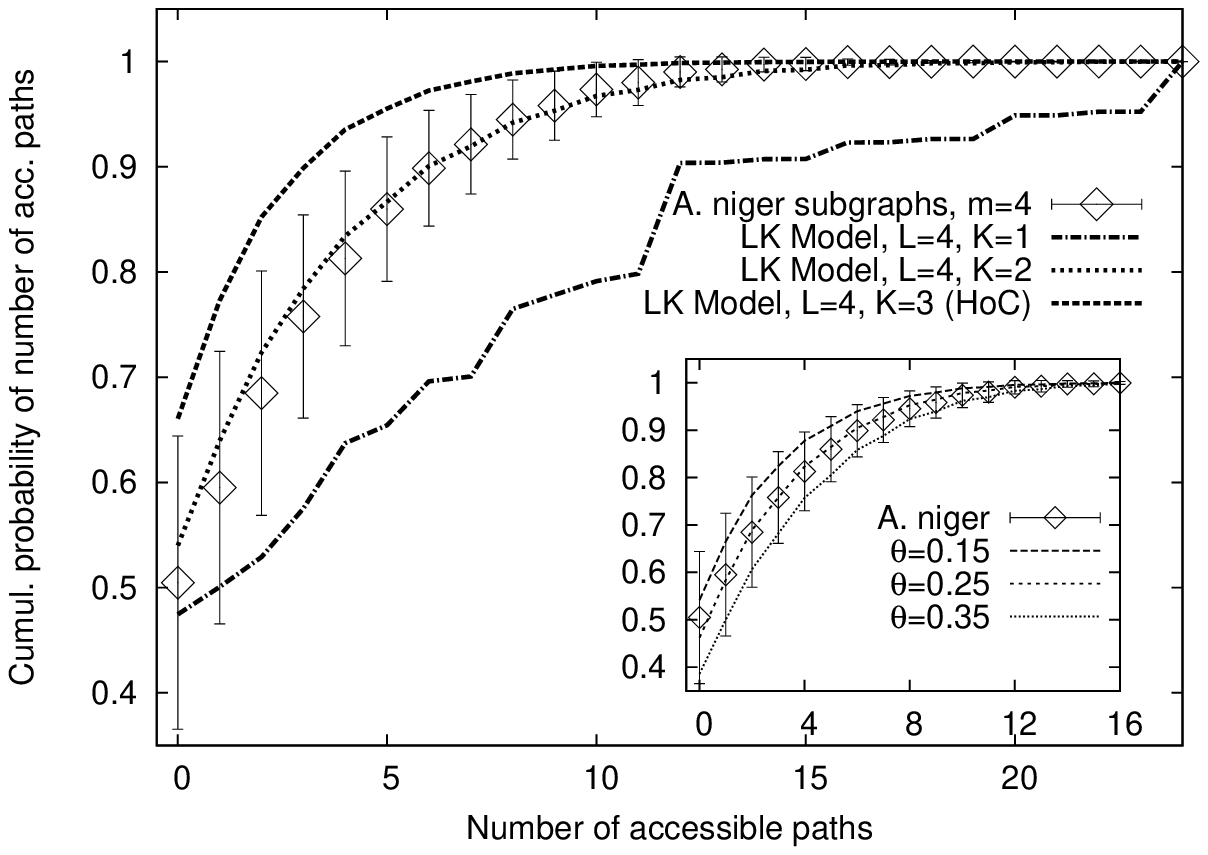}
\end{center}
\caption{{\bf Comparison of models to empirical data.} (A) Mean number of accessible paths 
for HoC, RMF and LK models compared to the empirical \emph{A. niger} data. With the
exception of the HoC model, all curves show an
increase of $\langle n_{L} \rangle$ with $L$. 
Both RMF (inset) and LK (main plot) models can be fit to the empirical data. Error
bars on the empirical data represent standard deviations obtained from the resampling analysis.
(B) Cumulative probability of the number of
    accessible paths as observed in the empirical fitness landscape
    compared to $LK$ (main plot) and RMF (inset) model. Error bars
    represent the standard deviation estimated by the resampling method.}
\label{en}
\end{figure}

Results of the subgraph analysis are displayed in Table~\ref{Table1} 
and in Fig.~\ref{en}. The data in Fig.~\ref{en}(a) show a systematic increase of the average
number of accessible paths with the mutational distance $m$ in the empirical data, which
rules out the null hypothesis of uncorrelated fitness values 
and is quantitatively consistent with the RMF model with $\theta
\approx 0.25$ (inset). The data for even subgraph sizes $m=2,4,6$ are equally
well described by the $LK$-model with $L=m$ and $K = m/2$ (main figure).
Alternatively, the empirical data can be compared to the results of a subgraph analysis of a
$LK$ fitness landscape with fixed $K$ and $L=8$ (Fig.~S5). While the fit between model and data is 
less satisfactory than that shown in Fig.~\ref{en}(a), the comparison is consistent with a value of 
$K$ between 4 and 5, which again indicates that each locus interacts with roughly half of 
the other loci. 

Further analysis of statistical properties of the 
\textit{A. niger} landscape confirms this conclusion. 
As an example, in Fig.~\ref{en}(b) we 
display the cumulative distribution of
the number of accessible paths 
\begin{equation}
\label{cumulative}
q_m(n) = \sum_{k=0}^n p_m(k)
\end{equation}
obtained from the analysis of the largest subgraph
ensemble with $m=4$. The main figure shows that good quantitative
agreement is achieved with the $L=4, K=2$ Kauffman model. The inset
displays a similar comparison to the RMF-model, which leads to the
estimate $\theta = 0.25 \pm 0.1$ for the roughness parameter, in close
agreement with the estimate obtained from $\langle n_m \rangle$.   

For the $m=4$ subgraph ensemble, the probability $p_4(0)$ of finding no accessible
path is approximately 0.5. Corresponding estimates $p_m(0)$ for other values of $m$
can be found in the last column of Table~\ref{Table1}. Up to $m=6$, the probability 
is found to increase with $m$, which implies that the ultimate increase of accessibility
(decrease in $p_m(0)$) predicted by the models cannot yet be seen on the scale of 
the empirical data. This is consistent with the estimates of the epistasis parameters
$\theta$ and $K$ mentioned above, for which the maximum in $p_L(0)$ is reached
at or beyond six loci (compare to Fig.~\ref{rmf_pl0}).

\section*{Discussion}

\subsection*{Evolutionary accessibility}

The models considered here represent a wide variety of intuitions
about 
fitness landscapes, from the null hypothesis of
uncorrelated fitness values through explicitly epistatic
models to the holey fitness landscapes derived from neutral
theory, thus covering all classes of fitness landscapes that are
expected to be relevant for real organisms.
With the exception of the extreme case of uncorrelated fitness values,
which is ruled out by comparison to the empirical data, 
all models show that fitness landscapes become highly accessible 
in the biologically relevant limit of large $L$: 
The probability of finding at least one accessible path is an
increasing function of $L$ which we conjecture to reach unity for $L \to
\infty$, and the expected number of paths grows with $L$ without
bounds. The latter feature limits
the repeatability of evolutionary trajectories.
 
In view of the robustness of these properties, we believe that their
origin lies in the topological structure of the configuration space:
The probability of accessibility of a given path (and thus the relative 
\textit{fraction} of accessible paths) decreases
exponentially with $L$, but this is overwhelmed by the combinatorial
proliferation of possible paths ($\sim L!$), see Eq.~(\ref{percon}) for
the neutral model and Eq.~(\ref{pi_Gumbel}) for the RMF model.
As we have imposed severe constraints on the adaptive process by 
prohibiting the crossing of fitness valleys by double mutations and
by only considering shortest paths, 
our estimate of accessibility is rather conservative. 
We therefore expect that naturally occurring, genome-wide fitness
landscapes should show a very high degree of accessibility as
well. 

A second general conclusion of our study is that
pathway accessibility in epistatic fitness landscapes is subject to
large fluctuations, as evidenced by the typical form of the
probability distribution $p_L(n)$ in Fig.\ref{pnl} and Figs.~S6, S7. 
For landscape dimensionalities $L$ in the range relevant for the available empirical 
studies, a substantial fraction of landscapes, given by $p_L(0)$, does not 
posses a single accessible pathway. On the other hand, for all models except the 
HoC model, the average number of
accessible pathways exceeds unity and increases rapidly
with increasing $L$. This implies that in those landscapes in which the maximum is accessible 
at all, it is typically accessible through a large number of pathways. 
For example, among the 70 $m=4$ subgraphs of the
\textit{A. niger} landscape, half do not contain a single accessible
path, but the average number of paths among the graphs with $n \geq 1$
is 4, and two subgraphs display as many as 10 accessible paths. 

This observation becomes relevant when applying similar analyses
to empirical fitness landscapes based on mutations that are collectively beneficial, 
such as the examples described in \cite{L2009,Chou2011,Khan2011}.  In these cases the adapted multiple 
mutant could not have been formed easily by
natural selection (alone) unless at least one selectively accessible pathway from the wildtype to the
mutant existed. The statistics of such landscapes is therefore biased towards larger
accessibility, and a comparison with random models should then be based on 
the probability distribution $p_L(n)$ conditioned on $n \geq 1$. The
general question as to whether landscapes formed by combinations of
beneficial or deleterious mutations have similar
topographical properties can only be answered by further empirical
studies. 

\subsection*{The \textit{A. niger} landscape}

The analysis of accessible mutational pathways in the empirical \textit{A. niger} data set
has allowed us to quantify the amount of sign epistasis in this landscape in terms of model
parameters like the roughness scale $\theta$ in the RMF model or the number of interacting
loci $K$ in the $LK$-model. Similar to a recent experimental study of viral adaptation \cite{Miller2011},
we ruled out the null model of a completely uncorrelated fitness landscape. Nevertheless our results
suggest that the epistatic interactions in this system are remarkably
strong.
To put our estimate of $K$ into perspective, we carried out
a subgraph analysis of the TEM $\beta$-lactamase antibiotic resistance landscape obtained in \cite{wddh2006} 
(Fig.~S8). In this case the number of loci is $L=5$, and the comparison of the mean number of accessible paths in 
subgraphs of sizes $m=2-4$ with simulation results for the $LK$-model suggests that $K \approx 1-2$, significantly
smaller than the estimate $K \approx L/2$ obtained for the \textit{A. niger} landscape. A low value of $K \leq 1$
was also found in the analysis of a DNA-protein affinity landscape for the set of all possible 10 base oligomers
\cite{Rowe2009}. 

Our finding of a high level of intergenic sign epistasis, compared to the examples of intragenic
epistasis considered in \cite{wddh2006} and \cite{Rowe2009}, contradicts the general expectation that epistatic interactions
should be stronger within genes than between genes \cite{Chou2011,Khan2011,Watson2010}. Note, however, that the comparisons among the available epistasis data are confounded by differences in the combined fitness of the
mutations involved: while the \textit{A. niger} mutations were chosen without a priori knowledge of their (combined) fitness effects, the
mutations considered in most studies were known to be collectively beneficial \cite{wddh2006,L2009,Lunzer2005,DaSilva2010,Chou2011,Khan2011}, and hence biased against negative epistatic combinations.

\subsection*{Population dynamics}

In the present paper we have focused on the existence of accessible 
mutational pathways, without explicitly addressing the probability that 
a given pathway will actually be found under a specific 
evolutionary scenario. This probability is expected to depend on population parameters, 
primarily on the mutation supply rate $N u$, in a complex way. In the SSWM 
regime characterized by $N u \ll 1$ it is straightforward in principle to assign probabilistic weights to mutational
pathways in terms of the known transition probababilities of the individual steps \cite{wddh2006,Orr2002}.
For larger populations additional effects come into play, whose bearing on accessibility and
predictability is difficult to assess. 

On the one hand, an increase in the mutation supply rate $Nu$ may bias adaptation towards the use of
mutations of large beneficial effects, which makes the evolutionary process more deterministic \cite{Jain2007} but
also more prone to trapping at local fitness maxima \cite{Jain2011}. While this reduces the accessibility of the
global optimum, at the same time the crossing of fitness valleys becomes more likely due to the fixation
of multiple mutations at once \cite{Weissman2009}, which tests mutants for their short-term evolvability 
\cite{Woods2011} and enlarges the set of possible mutational pathways.
We plan to address the interplay between landscape structure and population parameters in their effect on pathway accessibility 
in a future publication.

\section*{Materials and Methods}

\subsection*{Numerical Simulations}
For the numerical simulations of random
landscapes, fitness values were assigned to each of the $2^L$ genotypes
according to the ensemble to be sampled from (HoC, RMF or $LK$ model).
The number of paths was then found by a depth-first backtracking
algorithm implemented as an iterative subroutine starting at
the antipodal genotype and either moving ’forward’, i.e. towards the
global fitness maximum, or, if a local maximum is reached, going
back to the last genotype encountered before the local maximum. For
finding the probability $p_{L}(0)$ of no accessible paths, the search was
ended upon finding the first path, making this search much faster than
that for the full distribution of paths and thus enabling us to consider
much larger genotype spaces. Results were typically averaged over
$10^5$ realizations of the random landscape. In analyzing the empirical
\textit{A. niger} data, the same routines were used but with the measured fitness
values as input instead of fitness values sampled from one of the
models.

\subsection*{Analytic results for the RMF model}
It was argued above that both the expected number of
accessible paths $\langle n_{L}\rangle$ and the probability of no
accessible path $p_{L}(0)$ behave fundamentally different for $\theta =
0$ (HoC-model) and the RMF model with strictly positive $\theta$, even if
$\theta \ll 1$. Here we provide additional information on the relation (\ref{meanRMF}) 
and lend support to the statement that
typically $\pi_{L}$, the \textit{probability} of a given path being accessible,
decays exponentially in $L$.
Since by linearity of the expected
value $\langle n_{L}\rangle=L! \pi_{L}$, it is sufficient to consider 
$\pi_{L}$ to compute $\langle n_{L}\rangle$.

It was shown in \cite{fwk2010} that
\begin{equation}
\label{pi}
  \pi_{L}(\theta)\approx \frac{1}{L!} +\frac{\theta}{(L-2)!}\int \mathrm{d}x \; f^2(x) +\mathcal{O}(\theta^2)
\end{equation}
for $\theta \ll 1$, where $f(x)$ is the probability density of the random fitness contribution $x_{\boldsymbol{\sigma}}$. 
From this form it is clear that the HoC case $\theta = 0$
is quite different from the general case $\theta >0$.
Note that according to (\ref{pi}),  $\pi_{L}(\theta)$ still decays factorially as
$L \to \infty$. This changes, however, when higher order terms 
in $\theta$ are taken into account. 

For the special case when the fitness random contributions
are drawn from 
the Gumbel distribution $f(x)=\exp\left(-e^{-x}-x\right)$, the probability $\pi_L$ can be computed
explicitly for any $\theta$ \cite{fwk2010}. One obtains the expression 
\begin{equation}
\label{pi_Gumbel}
  \pi_{L}(\theta)=\frac{(1-e^{-c})^L}{\prod_{n=1}^L(1-e^{-cn})}
\end{equation}
with $c = (\pi/\sqrt{6}) \; \theta$.  
For large $L$, the denominator approaches a constant given by
\begin{equation}
  \lim_{L\to
    \infty}\prod_{n=1}^{L}(1-e^{-cn})\approx\sqrt{\frac{2\pi}{c}}\exp\left(-\frac{\pi}{6c}+\frac{c}{24}\right),
\end{equation}
and thus $\pi_L$ decays exponentially, $\pi_L \sim (1 - e^{-c})^L$. 
We expect this behavior to be generic for most choices of $f(x)$.

\subsection*{Data Set}
The fitness values constituting the 8-locus empirical data set 
are presented in Table S1. Here we briefly describe how these values were
obtained. A detailed description of the construction and
fitness measurement of the \textit{A. niger} strains is given elsewhere
\cite{dhv1997,pdk2009}.

Briefly, \emph{A. niger} is an asexual filamentous fungus with a predominantly
haploid life cycle. However, at a low rate haploid nuclei fuse
and become diploid; these diploid nuclei are often unstable and generate
haploid nuclei by random chromosome segregation. This alternation
of ploidy levels resembles the sexual life cycle of haploid
organisms and is termed parasexual cycle, since it does not involve
two sexes. We exploited the parasexual cycle of \textit{A.niger} to isolate
haploid segregants from a diploid strain that originated from a heterokaryon
between two strains that were isogenic, except for the presence
of eight phenotypic marker mutations in one strain, one on each
of its eight chromosomes. These mutations include, in increasing
chromosomal order, \emph{fwnA1} (fawn-colored conidiospores), \emph{argH12}
(arginine deficiency), \emph{pyrA5} (pyrimidine deficiency),
\emph{leuA1} (leucine
deficiency), \emph{pheA1} (phenyl-alanine deficiency), \emph{lysD25}
(lysine deficiency), 
\emph{oliC2} (oligomycin resistance), and \emph{crnB12} (chlorate resistance).
The wild-type strain only carried a spore-color marker (\emph{olvA1},
causing olive-colored conidiospores) on its first chromosome to allow
haploid segregants to be distinguished from the diploid mycelium
with black-colored conidiospores. Because these mutations were individually
induced with a low dose of UV and combined using the
parasexual cycle it was unlikely that the two strains differed at loci
other than those of the eight markers. 

From the $2^8 = 256$ possible
haploid segregants, 186 were isolated after forced haploidization
of the heterozygous diploid strain on benomyl medium from among
2,500 strains tested. Fitness of all strains was measured with two-fold
replication by measuring the linear mycelium growth rate in two perpendicular
directions during radial colony growth on supplemented
medium that allowed the growth of all strains, and was expressed relative
to the mycelium growth rate of the \emph{olvA1} strain with the highest
growth rate (see Table S1). As will be explained in the next section, missing genotypes are
assigned zero fitness. 

\subsection*{Data Analysis}
To analyze the data set, first one has to address the problem of
missing strains.  In the experiments, $186$ out of $256$ possible strains
were found
in approximately $2500$ segregants. Assume first that all genotypes
are equally likely to be found in the sample. Denoting the number
of segregants by $S$, the probability for a given strain to be missed by
chance is $p = (1 - 1/256)^S \approx 5.6 \times 10^{-5}$. The probability $p_n$ for
at most $n$ genotypes to have been missed is then given by a Poisson
distribution with mean $256 \times p \approx 0.014$. This gives the estimates
$p_0 \approx 1 - 256\times p = 0.986$ and $p_1 \approx 1 - (256\times
p)^2 \approx 0.9998$. For a more 
conservative estimate, one may assume that different genotypes have
different likelihoods to be found, which are uniformly distributed in
the interval $[r, 1]$ with $0 < r < 1$. Choosing $r = 0.274$ which
corresponds to the lowest relative fitness that was observed among
the viable genotypes, simulations of this scenario yield $p_0 \approx 0.74$
and $p_1 \approx 0.956$. We conclude that it is unlikely that more
than one viable genotype has been missed by chance. This justifies the assignement 
of zero fitness to the missing 70 genotypes.

Next we need to verify that accessibility in the empirical fitness
landscape is predominantly determined by sign epistasis among viable
genotypes, rather than by the presence of lethals.
As described in the main text, we consider subgraphs of
the \emph{A. niger} data set containing all combinations of $m$ of the eight
mutations in total. The set of subgraphs of size $m$ is composed of
${8 \choose m}$ distinct $m$-locus landscapes, each of which spans a
region in 
genotype space ranging from the wild type genotype shared by
all subgraphs to one particular $m$-fold mutant. We focus here on the
ensembles with $2 \leq m \leq 6$.

Key properties of the subgraph ensembles are summarized in Table~\ref{Table1}. 
The first column shows the total number ${8 \choose m}$
of subgraphs, and
the second column shows the number of viable subgraphs (VSG's),
defined as subgraphs which contain no non-viable strains. Two of the
four VSG's with $m = 5$ were previously analyzed in \cite{pdk2009}, and three of the 19 
VSG's with $m=4$ are shown in Fig.~1 of the main text.
To assess the impact of lethal genotypes on accessibility, let
$\langle n_{m} \rangle_\mathrm{leth}$ denote the average number of accessible
paths per subgraph (averaged across all subgraphs of fixed $m$) that would be present
if \emph{only} lethal states were allowed to block a path and the actual fitness values
of viable genotypes were ignored. Similarly, $\langle n_{m} \rangle$ denotes the
average number of accessible paths per subgraph for fixed $m$ if both
mechanisms for blocking are taken into account. 
Comparison between the two numbers, displayed in the fourth and fifth column of 
Table~\ref{Table1}, shows that the
contribution of the lethal mutants to reducing pathway accessibility is
relatively minor. For example, for $m = 4$ lethals reduce the number
of accessible paths from $4! = 24$ to $12$, by a factor of $0.5$,
whereas the epistasis among viable genotypes leads to a much more substantial
further reduction from $12$ to $2$, by a factor of $1/6$; for $m=6$ the
corresponding factors are $0.34$ and $0.008$. We conclude that pathway
accessibility is determined primarily by epistasis among viable
genotypes.

Inspection of the VSG's shows that the role of different mutations in causing
lethality is strikingly inhomogeneous. In particular, we find that the
lysine deficiency mutation \emph{lysD25} is not present in any of
VSG's, whereas the distribution of the other mutations across the VSG's
is roughly homogeneous. 
The \emph{lys} mutation is also strongly overrepresented in the
non-viable strains, being present in $62$ out of $70$ cases.
The main features of the set of lethal mutations can be captured
in a simple model in which the presence of a mutation $i$ leads to a
non-viable strain with probability $q_i$, and different mutations interact
multiplicatively, such that a strain containing two mutations $i$ and $j$ is
viable with probability $(1 - q_i)(1 - q_j)$. The data for the number
of VSG's for different $m$ cannot be described assuming the $q_i$ to
be the same for all mutations, but a two-parameter model assigning
probability $q_{lys}$ to the $lys$ mutation and a common value $q_0
\ll q_{lys}$ to
all others suffices. Simple analysis show that under this model the expected
total number of viable strains is $N_\mathrm{viable}=(2 -  q_{lys})(2 -  q_0)^7$,
while the total number of viable strains in the subset of strains excluding
\textit{lys} is $\tilde{N}_\mathrm{viable}=(2 - q_0)^7$. With $N_\mathrm{viable}=186$ and
$\tilde{N}_\mathrm{viable}=120$ we obtain the estimates $q_{lys} \approx
0.45$ and $q_0 \approx 0.018$. 
Given that the VSG's do not contain the \textit{lys} mutation, the expected
number of VSG's depends only on $q_0$, and is given by
\begin{equation}
\label{Cm}
C_{m}={L \choose m}(1-q_0)^{m2^{m-1}}.
\end{equation}
The results for the expected number of viable subgraphs are shown in brackets
in the third column of Table~\ref{Table1}, and are seen to match the data very
well. Similarly, the expected number of paths that do not contain any lethal
genotypes can be computed analytically, resulting in the expression 
\begin{equation}
\label{nleth2}
\langle n_m \rangle_{\mathrm{leth}} = m! \; (1-q_0)^{\frac{m(m+1)}{2}}
\left\{ 1 - \frac{m}{L} + \frac{1}{L} \frac{1 - q_{lys}}{q_{lys} - q_0}
\left[1 - \left(\frac{1 - q_{lys}}{1 - q_0} \right)^m \right] \right\},
\end{equation}
which is shown in brackets in the fourth column of Table~\ref{Table1}.

\subsection*{Resampling procedure}

The accessibility
of mutational pathways in the \textit{A. niger} data set was analyzed
using two different approaches. The first approach is based on a
single set of fitness values obtained by averaging the two replicate
fitness measurements for each strain; these average fitness values are shown in Table S1. 
In the second approach
the influence of errors in the fitness measurements was
taken into account by using a resampling procedure previously
described in \cite{pdk2009}. In this approach the fitness assigned
to each viable genotype is a normally distributed random variable with the mean given
by the average of the two fitness measurements and a common standard
deviation $s_0 \approx 0.03$
estimated from the mean squared differences between replicate fitness
values in the entire data set; the fitness of genotypes identified as
non-viable remains zero. Statistical properties of accessible pathways
are then computed by averaging over $10^5$ realizations of this
resampled landscape ensemble. Empirical data points and error bars shown in
Fig.~\ref{en} 
represent the mean and standard deviations obtained from
the second approach. Results obtained by directly analyzing the mean fitness landscape
(first approach) do not differ significantly from those presented here.

\section*{Acknowledgments}
We thank Simon Gravel, Su-Chan Park, Chris Marx, Martijn Schenk and Shamil Sunyaev
    for useful discussions and suggestions. 

\bibliography{franke}


\section*{Tables}

\begin{table}[tbh]
\caption{{\bf Subgraphs of the \textit{A. niger} data set} }

\begin{tabular}{lcccccc} \\ \hline
$m$ & \# SG & \# VSG & $\langle n_m \rangle_\mathrm{leth}$ & $\langle n_m \rangle$ & $p_m(0)$   \\ \hline
2 & 28 & 20 (19.5) &  1.61 (1.72) & 0.82  & 0.36  \\ 
3 & 56 & 29 (28.1) &  4.05 (4.22) & 1.34  & 0.39   \\ 
4 & 70 & 19 (19.5) &  12.53 (13.19) & 2.01 & 0.50     \\ 
5 & 56 & 4  (4.9) &  55.32 (48.81) &3.16  & 0.63  \\
6 & 28 & 0  (0.2) & 246.0 (201.16) & 6.07  & 0.68 \\ \hline  
\end{tabular}
\begin{flushleft}The table summarizes properties of subgraphs of sizes $m=2,...,6$ of the empirical
\textit{A. niger} fitness landscape. Second column shows the total number of subgraphs
${8 \choose m}$ and third column the number of viable subgraphs not containing any non-viable
genotypes, with the model prediction (\ref{Cm}) given in brackets. Fourth column contains the 
number of accessible paths that would be present if accessibility were reduced only because
of the presence of non-viable genotype, with the model prediction (\ref{nleth2}) shown in brackets.
Finally, the last two columns show the mean number of accessible paths $\langle n_m \rangle$ 
and the probability of no accessible path $p_m(0)$, respectively, computed from the full subgraph
ensemble. 
\end{flushleft}
\label{Table1}
\end{table}

\newpage

\section*{Figures}
 
\end{document}